\documentclass{aa}
\usepackage{astron,times,graphics,latexsym}
\begin{document}

\thesaurus{04 (02.14.1; 08.01.1; 08.03.1; 08.23.2; 10.01.1; 10.05.1)}
\title{The origin of carbon, investigated by spectral analysis of solar-type
stars in the Galactic Disk$^{\star}$}
\author{B. Gustafsson
\and T. Karlsson
\and E. Olsson
\and B. Edvardsson
\and N. Ryde}
\offprints{B. Gustafsson (Bengt.Gustafsson@astro.uu.se)\\
$^{\star}$ Based on observations carried out at the European Southern Observatory, La Silla, Chile.}
\institute{Uppsala Astronomical Observatory, Box 515, SE-751 20, Uppsala, 
  Sweden}
\date{Received 31 July 1998 / Accepted 21 October 1998}
\titlerunning{The origin of carbon}
\maketitle
\begin{abstract}
Abundance analysis of carbon has been performed in a sample of 80 late F and
early G type dwarf stars in the metallicity range
$\mathrm{-1.06\le[Fe/H]\le0.26}$ using the forbidden [C {\sc i}] line at
$8727\mbox{{\AA}}$. This line is presumably less 
sensitive to temperature, atmospheric structure and departures from LTE
than alternative carbon criteria. We find that $\mathrm{[C/Fe]}$ decreases 
slowly with increasing $\mathrm{[Fe/H]}$ with an overall slope of 
$-0.17\pm0.03$.
Our results are consistent with carbon enrichment by superwinds of metal-rich 
massive stars but inconsistent with a main origin of carbon in low-mass stars. This follows in particular from a comparison 
between the relation of $\mathrm{[C/O]}$ with metallicity for the Galactic stars and the corresponding relation observed for dwarf irregular galaxies. The 
significance of intermediate-mass stars for the production of carbon in
the Galaxy is still somewhat unclear.

\keywords{Nuclear reactions, nucleosynthesis, abundances --
          Stars: abundances -- Stars: carbon -- Stars: Wolf-Rayet -- 
          Galaxy: abundances -- Galaxy: evolution}
\end{abstract}

\section{Introduction}

The stellar
origin of carbon became clear with the discovery of the
Triple Alpha reaction (\"Opik 1951\nocite{opik}; Salpeter
1952\nocite{sal115}) and the beautiful demonstration by Hoyle \cite*{hoyle}
that the high carbon abundance would require a resonance state  at
$7.65\mbox{MeV}$ of carbon, a prediction that was soon verified experimentally.
But in which type of stars was carbon formed? 

In their classical paper Burbidge et al. \cite*{b2fh} 
suggested carbon to be provided by mass loss from red giants and supergiants. 
Arnett \& Schramm \cite*{arnett}
argued that massive stars might be most effective. They
discussed the composition of matter estimated to be 
ejected from stars (with helium core masses above $4 \mathcal{M}_{\odot}$,
i.e.  $\mathrm {\mathcal{M}_{tot} \ge 10 \mathcal{M}_{\odot}}$) and found 
$\mathrm{C/O}$ ratios consistent with the solar system
values. The yield prescriptions developed by Talbot \& Arnett (1973, 1974)
\nocite{t1} \nocite{t2}
consequently ascribed carbon production to the massive stars. However, 
Truran \cite*{truran} and Tinsley \cite*{tinsley} argued that 
carbon stars of lower mass could also be significant sites for 
carbon. Dearborn et al. \cite*{dea} also suggested that low mass
stars (as evidenced by carbon stars  and carbon-rich planetary nebulae) may be
a significant source of ${\rm \,^{12}C}$, while Iben \& Truran \cite*{iben78}
concluded from thermally pulsing models that
intermediate mass stars and high mass stars contributed carbon in roughly
equal amounts. The arguments for significant contributions of carbon by 
low-mass and intermediate-mass stars were further elaborated
by Tinsley \cite*{tinIAU} and  Sarmiento \& Peimbert \cite*{sarmiento}.
Timmes et al. \cite*{timmes} in their 
models of Galactic chemical evolution with then available yields for
supernovae as well as for low- and intermediate mass stars, found
very significant contributions of carbon from the latter to the Galactic
Disk. (For a general review on the significance of low- and intermediate-mass
carbon stars for Galactic nucleosynthesis, see
Gustafsson \& Ryde (1998\nocite{bg_ryde}).) Recently, Kobulnicky \& Skillman 
\cite*{kob_skill} concluded from the correlation between $\mathrm{C/O}$
and  $\mathrm{N/O}$ ratios found in H {\sc ii} regions in three starforming
galaxies, that the formation of N and C is coupled and presumably occurs in 
intermediate or low mass stars.  
Prantzos et al. \cite*{prantzos}, however, adopting the idea of
Maeder \cite*{maeder} that radiatively driven
massive winds from high-mass stars should provide huge amounts of helium 
and carbon, found these stars to be the main contributors while the role of 
intermediate-mass stars seemed much less significant.
Similarly, Garnett et al. \cite*{garnett95} observed that the C/O ratio 
increased with increasing O/H for dwarf irregular galaxies, a trend which they 
found to be consistent with the suggestion that carbon is produced in massive
stars with yields dependent on metallicity, as is expected for stars with
radiatively driven winds.
\par
The shifting views as regards the relative role of high-mass, 
intermediate-mass and low-mass stars have reflected the different
uncertainties and conjectures concerning the carbon production, dredge-up and 
mass loss from stars of different mass and metallicity. 
Although considerable progress has been made in these respects, 
there are still very severe remaining uncertainties. 
Also as regards observed carbon abundances for stars and planetary 
nebulae the uncertainties have been, and are still, considerable which
makes checks of the theoretical results, as well as more direct empirical
approaches, difficult. We shall here present an attempt to improve in this
latter respect, as regards carbon abundances for solar-type stars in the
Galactic Disk. The idea is to use high S/N observations of 
a hitherto little used forbidden carbon
line in the infrared as the single criterion. The strength of this line, 
in comparison with other carbon abundance criteria,
is not at all as sensitive to other parameters as to the carbon abundance.  
We shall also see that with high quality data it is 
possible to carry out more direct tests concerning the role of stars of
different masses; tests that are independent on the results of the
calculations of yields.   

\section{Observations and data reductions}

The observations were made at European Southern Observatory (ESO) with the 
$1.4\mbox{m}$ Coud\'e Auxillary Telescope (CAT) coupled to the Coud\'e Echelle
Spectrometer (CES) and the Long Camera during 3 observing runs with the
following CCD detectors: July 2-15 1994 CCD\#30, Dec 29 1994-January 4 1995
CCD\#34 and Sept 14-26 1995 CCD\#34. The slit-viewing auto-guiding system
ensured efficient light collection. The focusing and alignment of the CCD
cameras were checked before each observing night.
\par
Spectra centered near the $8727\mbox{{\AA}}$ [C {\sc i}] line and with a
wavelength range of about $70\mbox{{\AA}}$ and a resolution of 65,000 
were obtained for 87 southern stars of the Edvardsson et al. \cite*{edvardsson}
(cited as EAGLNT in the following) survey with different metallicities. 
Depending on the stellar magnitudes and weather conditions integrations of
between 20 seconds (Procyon, $V=0.37$) and 8 times 60 minutes (HD 199289,
$V=8.29$) were obtained for the stars. The long integration times were motivated
by the need to obtain high S/N ratios (preferably above 200) 
for the generally very weak [C {\sc i}] line.
In order to increase the S/N ratio, slightly
different central wavelengths were chosen for separate observations of the same
star. A typical S/N ratio for the final, summed up spectra was generally between
200 and 300. No spectrum had a S/N ratio below 150.
\par
The data reductions were performed either on line with the ESO IHAP software or
later with the ESO MIDAS package, and contained the following steps:
\begin{itemize}
\item Subtraction of bias and dark signal from the stellar frames and of bias
from the flat-field frames.
\item Division of the stellar frame by a flat-field frame obtained with
unaltered set-up, and within hours of the stellar frame.
\item Summation of the relevant CCD rows to produce a one-dimensional spectrum.
\item Wavelength calibration by means of a Th/Ar lamp spectrum obtained in
connection with the stellar frame.
\item A 3rd order spline function was fitted to the pre-defined continuum points
in the spectrum, and finally the spectrum was divided by this function to yield
a relative flux scale with continuum level of $1.00$.
\end{itemize}
\par
For stars where more than one spectrum of the region was obtained, both the
individual spectra and also the rebinned and summed total spectrum were 
alternatively used in the data analysis.
\begin{figure}
 \resizebox{\hsize}{!}{\includegraphics{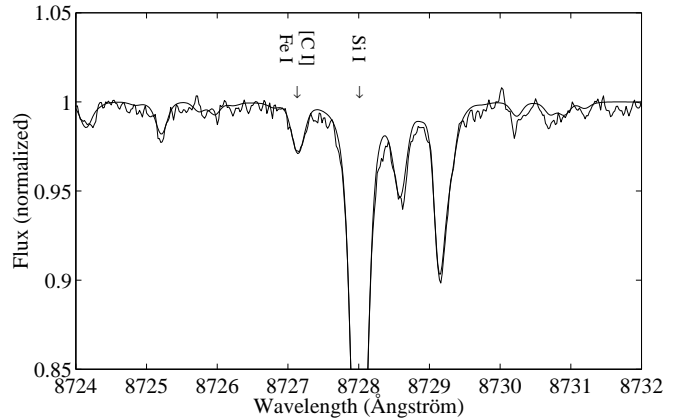}}
 \caption{The observed and synthetic spectra of HR 1294 (HD 26491) with the
carbon, iron and silicon lines indicated}
 \label{specex}
\end{figure} 
\section{Analysis}
\subsection{Model atmospheres}
The stellar spectra were synthesised with the Uppsala spectrum synthesis
program, using the opacity-sampling MARCS model atmospheres described in EAGLNT.
These are line-blanketed, one-dimensional atmospheres with mixing-length 
convection, produced under the assumption of LTE.
An example of a comparison between observed and synthetic spectra is shown in 
Fig. 1.

\subsection{Atmospheric parameters}
Originally, the atmospheric parameters $T_{\mathrm{eff}}$, $\log g$ and 
$\mathrm{[Fe/H]}$ of EAGLNT were adopted, and $\xi_{\mathrm{t}}$ was calculated
with the formula given in EAGLNT
\begin{equation}
\xi_{\mathrm{t}}=1.25+8\cdot 10^{-4}(T_{\mathrm{eff}}-6000)-
1.3(\log g-4.5)\mbox{km\,s$^{-1}$}.
\end{equation}
In EAGLNT, the surface gravities ($g_{\mathrm{phot}}$) were based on the
$c_{\mathrm{1}}$ index of the Str\"omgren photometry. However, surface gravities
were alternatively estimated for the stars from the parallaxes given in the
HIPPARCOS catalogue \cite{hipparcos} using the equation
\begin{eqnarray}
\label{eq:logg}
\log g_{\pi}=4\log \frac{T_{\mathrm{eff}}}{T_{\mathrm{eff}}^{\odot}}+\log
\frac{\mathcal{M}}{\mathcal{M}_{\odot}}+2\log \pi +\nonumber\\
0.4(V+BC-0.26)+4.44.
\end{eqnarray}
\begin{figure}
 \resizebox{\hsize}{!}{\includegraphics{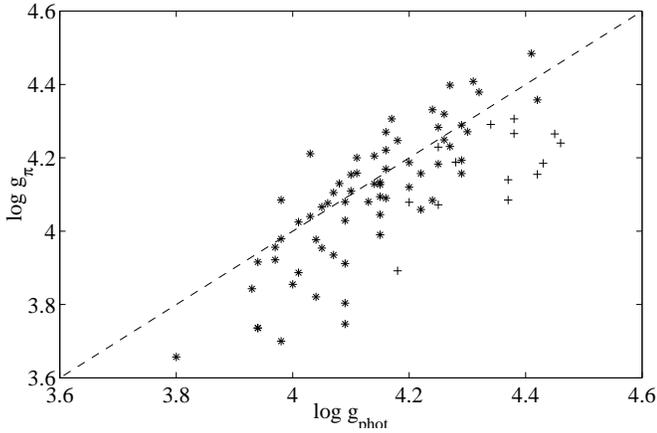}}
 \caption{Comparison between $\log g_{\mathrm{phot}}$ given in
Edvardsson et al. (1993) (EAGLNT) and $\log g_{\mathrm{\pi}}$ estimated from
Hipparcos parallaxes. The plusses represent stars with
$\mathrm{[Fe/H]}\leq -0.6$. The dashed line indicates a one-to-one relation to
guide the eye}
 \label{fig:logg}
\end{figure}
The masses for the stars were estimated from evolutionary tracks by
Schaller et al. \cite*{schaller} and Claret \& Gimenez \cite*{claret}. Due to
the uncertainty in the models and the few tabulated steps in $Z$, the
uncertainty in the mass estimate may be
$\Delta\mathcal{M}=0.2 \mathcal{M}_{\odot}$. Ng \& Bertelli \cite*{ng} have also
estimated masses for these stars showing a good agreement compared to our
values. The bolometric corrections were estimated according to
Allen \cite*{allen}. For our stars the bolometric corrections are small.
The V-magnitudes from the HIPPARCOS catalogue have been used.
\par 
A comparison between $g_{\mathrm{\pi}}$ and $g_{\mathrm{phot}}$ is shown in
Fig. \ref{fig:logg}. In Table \ref{data} the differences between
$\log g_{\mathrm{phot}}$ and $\log g_{\mathrm{\pi}}$ are given.
It is seen that $g_{\mathrm{phot}}$ is greater in most cases, the mean
difference is 0.06 dex, with a formal scatter of 0.12 dex.
In Fig. \ref{fig:logg} one can also see that all stars with
$\mathrm{[Fe/H]}\leq -0.6$ have a lower $g_{\mathrm{\pi}}$ than
$g_{\mathrm{phot}}$. The systematic trend that low $\log g$ values become even
lower when estimated from parallaxes might be due to errors in the calibration
of Str\"omgren photometry. 
\par
We have here chosen to subsequently use $\log g_{\mathrm{phot}}$ since the
statistical errors in $\log g_{\mathrm{\pi}}$ are greater. However, systematic
errors in the surface gravities may affect the carbon abundance determinations
and the comparison with $\log g_{\mathrm{\pi}}$ could give a hint as regards
the effects of systematic errors. This is discussed further below.

\subsection{Atomic and molecular line data}

The [C {\sc i}] $8727\mbox{{\AA}}$ line is a weak feature, located in the far
wing of a stronger Si {\sc i} line, and care must be exercized so that all
relevant blends affecting the line as well as the neighbouring continuum regions
are included in the synthetic spectra used in the analysis.  The line list used
contains all lines existing in the VALD data base (Piskunov et al. 1995) 
\nocite{nick} 
between $8705\mbox{{\AA}}$ and $8750\mbox{{\AA}}$. In addition, data for more
than one hundred CN lines, in the interval $8722\mbox{{\AA}}$ to
$8732\mbox{{\AA}}$, were added to the list by Bertrand Plez. 
Theoretically calculated wavelengths of molecular lines are difficult to obtain
to high accuracy. Laboratory measured wavelengths were found for three of the CN
lines in this interval while the rest were corrected in a semi-empirical way
using a program that Patrick de Laverny kindly made available to us. The three
laboratory measured wavelengths were also compared to the corrected wavelengths.
They agreed to within $0.04\mbox{{\AA}}$. There are evidently many CN lines in
the region but none of them seems to affect the [C {\sc i}] 8727 \AA\ line
significantly.    
The lines in the neighbourhood of the [C {\sc i}] line are listed in
Table \ref{linelist}.
\par
We also searched for telluric emission and absorption lines in the region of
interest. These lines can influence the effective equivalent width of the carbon
line if not properly removed.
We found no evidence for such undesired features.

\begin{table}
  \caption{Atomic and molecular line data for lines in the immediate neighbourhood
of the 8727 {\AA} [C {\sc i}] line. Only lines with an equivalent width
 $W_{\mathrm{\lambda}}>0.05\mbox{m{\AA}}$ are given. The first five columns give, respectively, 
the rest wavelength, the lower excitation energy, the logarithm of the product
of the oscillator strength and the statistical weight, a multiplicative factor
to the classical van der Waals damping constant and the radiation damping 
constant. $^{\star}$ indicates a change in $\lambda$, $\log gf$, $\Gamma_{6}$ and/or 
$\Gamma_{\mathrm{rad}}$ relative to the original value in VALD}
   \begin{tabular}{lllllll}
   \hline
  ~$\lambda$ & $\chi_{\mathrm{l}}$ & $\log gf$    & $\Gamma_{6}$ & $\Gamma_{\mathrm{rad}}$& &\\
  ~[{\AA}]   & [eV]       &             &               & [s$^{-1}$]&&    \\
  \\
  \hline  
  C {\sc i}  $\log \epsilon_{\odot}=8.55$	&&&&&&				\\
  $8727.14^{\star}$  & $1.264$      &$-8.240^{\star}$       & $2.50$          & $0.500\mathrm{E}+00^{\star}$&&\\
  
  N {\sc i}  $\log \epsilon_{\odot}=7.97$	&&&&&&				\\
  $8728.90$  &$10.330$      &$-1.164$       & $2.50$          & $4.169\mathrm{E}+08$&&       \\
  
  Si {\sc i} $\log \epsilon_{\odot}=7.55$     &&&&&&                       \\
  $8728.01$  & $6.181$      &$-0.370^{\star}$       & $2.00^{\star}$          & $1.194\mathrm{E}+07^{\star}$&&\\       
  $8728.59$  & $6.181$      &$-1.550^{\star}$       & $2.50$          & $1.888\mathrm{E}+07^{\star}$&&\\       
  $8742.45$  & $5.871$      &$-0.450$       & $1.30^{\star}$          & $6.839\mathrm{E}+07^{\star}$      &&\\
  
  Ca {\sc i} $\log \epsilon_{\odot}=6.36$      &&&&&&				\\
  $8725.21^{\star}$  & $4.624$      &$-0.910^{\star}$       & $2.50$          & $8.892\mathrm{E}+07$&&\\         
  $8725.05$  & $4.535$      &$-2.400$       & $2.50$          & $2.023\mathrm{E}+07$      &&\\   
  
  Ti {\sc i} $\log \epsilon_{\odot}=5.02$      &&&&&&				\\
  $8725.96^{\star}$  & $1.734$      &$-2.158^{\star}$       & $2.50$          & $6.577\mathrm{E}+07$&&\\

  Fe {\sc i} $\log \epsilon_{\odot}=7.50$	&&&&&&				\\
  $8727.13$  & $4.186$      &$-3.930$       & $2.50$          & $6.095\mathrm{E}+07$      &&\\            
  
  CN					&&&&&&						 \\
  $8726.23^{\star}$   & $1.198$     &$-1.788$       & $2.50$          & $1.000\mathrm{E}+05$ & &\\
  $8726.80$  & $1.335$      &$-1.481$       & $2.50$          & $1.000\mathrm{E}+05$       &&\\   
  $8727.81$  & $1.559$      &$-1.465$       & $2.50$          & $1.000\mathrm{E}+05$       &&\\    
\hline
\end{tabular} 
  \label{linelist}
\end{table}

\setcounter{table}{1}
 \begin{table*}
   \caption{Column 2-4 give the atmospheric parameters for the stars adopted from EAGLNT (Edvardsson et al. 1993) 
and column 5 contains our measured carbon abundances. Column 6 shows the oxygen abundance taken from EAGLNT (they do not give an oxygen abundance for all the stars in our sample) and column 7 is the logarithm of the age also adopted from EAGLNT. Column 8 and 9 contain data from the HIPPARCOS catalogue, the parallax in mas resp. the V-magnitude. Column 10 give the estimate of the mass from evolutionary tracks as described in the text and column 11 give the difference between the photometric $\log g$ adopted from EAGLNT and $\log g$ calculated using HIPPARCOS data}
\label{data}
   \begin{tabular}{l|ccr|r|rr|rrrr}
   \hline
   Star   & $T_{\mathrm{eff}}$  & $\log g$ & $\mathrm{[Fe/H]}$  & $\mathrm{[C/H]}$ & $\mathrm{[O/H]}$  &$\log \tau_9$   & $\pi$    & $V$     & $\mathcal{M}/\mathcal{M}_{\odot}$ & $\Delta\log g$\\
   Id.    & [K]   &[cgs]  &	    &       &        &[Gyr]     & mas      &mag       & 	&[cgs] \\ 		
   \hline			   	                     
   HR 33   &$ 6204$  &$ 4.07 $ &$-0.38 $   &$ -0.35 ~~$&$ -0.35 $ &$ 0.74 $    &$ 52.94 $   &$4.89  $ &$ 1.12 $&$ -0.04 $\\
   HR 35   &$ 6577$  &$ 4.26 $ &$-0.10 $   &$ 0.03  ~~$&$ -0.24 $ &$ 0.29 $    &$ 45.85 $   &$5.24  $ &$ 1.18 $&$ 0.01  $\\
   HR 107  &$ 6488$  &$ 4.08 $ &$-0.37 $   &$ -0.15 ~~$&$       $ &$ 0.53 $    &$ 27.51 $   &$6.05  $ &$ 1.25 $&$ -0.05 $\\
   HR 140  &$ 6408$  &$ 4.18 $ &$ 0.05 $   &$ 0.07  ~~$&$ -0.12 $ &$ 0.46 $    &$ 39.03 $   &$5.57  $ &$ 1.33 $&$ -0.07 $\\
   HR 235  &$ 6254$  &$ 4.32 $ &$-0.15 $   &$ -0.04 ~~$&$       $ &$ 0.49 $    &$ 64.69 $   &$5.17  $ &$ 1.05 $&$ -0.06 $\\
   HR 366  &$ 6474$  &$ 4.10 $ &$-0.32 $   &$ -0.24 ~~$&$       $ &$ 0.54 $    &$ 41.01 $   &$5.14  $ &$ 1.25 $&$ -0.01 $\\
   HR 368  &$ 6517$  &$ 4.01 $ &$-0.24 $   &$ -0.15 ~~$&$       $ &$ 0.49 $    &$ 22.79 $   &$5.70  $ &$ 1.41 $&$ 0.12  $\\ 
   HR 370  &$ 6081$  &$ 4.26 $ &$0.12  $   &$ 0.16  ~~$&$ 0.08  $ &$ 0.54 $    &$ 66.43 $   &$4.97  $ &$ 1.18 $&$ -0.06 $\\
   HR 573  &$ 6239$  &$ 4.25 $ &$-0.34 $   &$ -0.25 ~~$&$ -0.32 $ &$ 0.77 $    &$ 37.97 $   &$6.10  $ &$ 1.05 $&$ -0.03 $\\
   HR 740  &$ 6436$  &$ 3.94 $ &$-0.25 $   &$ -0.17 ~~$&$ -0.29 $ &$ 0.49 $    &$ 38.73 $   &$4.74  $ &$ 1.33 $&$ 0.02  $\\
   HR 1010 &$ 5889$  &$ 4.41 $ &$-0.23 $   &$ -0.20 ~~$&$ -0.27 $ &$ 0.94 $    &$ 82.79 $   &$5.24  $ &$ 1.00 $&$ -0.07 $\\
   HR 1083 &$ 6769$  &$ 4.10 $ &$-0.11 $   &$ -0.05 ~~$&$ -0.22 $ &$ 0.26 $    &$ 46.65 $   &$4.71  $ &$ 1.33 $&$ -0.05 $\\
   HR 1101 &$ 5981$  &$ 4.15 $ &$-0.11 $   &$ -0.02 ~~$&$ -0.09 $ &$ 0.77 $    &$ 72.89 $   &$4.29  $ &$ 1.05 $&$ 0.11  $\\
   HR 1173 &$ 6739$  &$ 4.11 $ &$0.09  $   &$ 0.20  ~~$&$ -0.06 $ &$ 0.22 $    &$ 55.79 $   &$4.22  $ &$ 1.50 $&$ -0.05 $\\
   HR 1257 &$ 6301$  &$ 3.97 $ &$0.04  $   &$ 0.10  ~~$&$       $ &$ 0.48 $    &$ 28.87 $   &$5.36  $ &$ 1.50 $&$ 0.05  $\\
   HR 1294 &$ 5732$  &$ 4.16 $ &$-0.18 $   &$ -0.14 ~~$&$ -0.19 $ &$ 1.03 $    &$ 43.12 $   &$6.37  $ &$ 0.90 $&$ -0.11 $\\
   HR 1536 &$ 5886$  &$ 3.98 $ &$0.14  $   &$ 0.15  ~~$&$ 0.06  $ &$ 0.64 $    &$ 37.73 $   &$5.77  $ &$ 1.18 $&$ -0.11 $\\
   HR 1545 &$ 6425$  &$ 4.11 $ &$-0.33 $   &$ -0.14 ~~$&$ -0.36 $ &$ 0.59 $    &$ 28.28 $   &$6.27  $ &$ 1.18 $&$ -0.09 $\\
   HR 1673 &$ 6442$  &$ 4.05 $ &$-0.30 $   &$ -0.17 ~~$&$ -0.31 $ &$ 0.55 $    &$ 39.99 $   &$5.11  $ &$ 1.25 $&$ -0.02 $\\
   HR 1687 &$ 6596$  &$ 4.15 $ &$0.26  $   &$ 0.33  ~~$&$       $ &$ 0.19 $    &$ 26.04 $   &$5.89  $ &$ 1.50 $&$ 0.02  $\\
   HR 1983 &$ 6398$  &$ 4.29 $ &$-0.07 $   &$ -0.01 ~~$&$ -0.14 $ &$ 0.43 $    &$111.49 $   &$3.59  $ &$ 1.12 $&$ 0.00  $\\
   HR 2233 &$ 6347$  &$ 4.07 $ &$-0.17 $   &$ -0.09 ~~$&$       $ &$ 0.55 $    &$ 28.02 $   &$5.62  $ &$ 1.25 $&$ 0.14  $\\
   HR 2493 &$ 6063$  &$ 4.24 $ &$-0.38 $   &$ -0.33 ~~$&$ -0.40 $ &$ 0.94 $    &$ 37.60 $   &$6.43  $ &$ 1.00 $&$ -0.09 $\\
   HR 2530 &$ 6595$  &$ 4.16 $ &$-0.43 $   &$ -0.21 ~~$&$ -0.41 $ &$ 0.42 $    &$ 33.45 $   &$5.78  $ &$ 1.25 $&$ -0.06 $\\
   HR 2548 &$ 6460$  &$ 4.06 $ &$-0.20 $   &$ -0.06 ~~$&$ -0.22 $ &$ 0.50 $    &$ 39.66 $   &$5.14  $ &$ 1.25 $&$ -0.02 $\\
   HR 2835 &$ 6184$  &$ 4.17 $ &$-0.55 $   &$ -0.39 ~~$&$       $ &$ 0.81 $    &$ 32.43 $   &$6.54  $ &$ 1.05 $&$ -0.14 $\\
   HR 2883 &$ 5976$  &$ 4.18 $ &$-0.75 $   &$ -0.47 ~~$&$ -0.49 $ &$ 1.03 $    &$ 33.40 $   &$5.90  $ &$ 0.80 $&$ 0.29  $\\
   HR 2906 &$ 6167$  &$ 4.09 $ &$-0.18 $   &$ -0.04 ~~$&$ -0.06 $ &$ 0.62 $    &$ 38.91 $   &$4.44  $ &$ 1.41 $&$ 0.34  $\\
   HR 2943 &$ 6704$  &$ 4.03 $ &$-0.02 $   &$ 0.10  ~~$&$ -0.05 $ &$ 0.22 $    &$285.93 $   &$0.40  $ &$ 1.50 $&$ -0.01 $\\
   HR 3018 &$ 5822$  &$ 4.42 $ &$-0.78 $   &$ -0.62 ~~$&$ -0.50 $ &$ 1.18 $    &$ 65.79 $   &$5.36  $ &$ 0.70 $&$ 0.27  $\\
   HR 3220 &$ 6536$  &$ 4.13 $ &$-0.26 $   &$ -0.18 ~~$&$ -0.22 $ &$ 0.49 $    &$ 46.75 $   &$4.74  $ &$ 1.25 $&$ 0.05  $\\
   HR 3578 &$ 5965$  &$ 4.37 $ &$-0.82 $   &$ -0.49 ~~$&$ -0.54 $ &$ 1.11 $    &$ 46.90 $   &$5.80  $ &$ 0.70 $&$ 0.29  $\\
   HR 4012 &$ 6124$  &$ 4.09 $ &$0.14  $   &$ 0.16  ~~$&$       $ &$ 0.59 $    &$ 19.27 $   &$6.02  $ &$ 1.58 $&$ 0.29  $\\
   HR 4039 &$ 6158$  &$ 4.30 $ &$-0.38 $   &$ -0.30 ~~$&$       $ &$ 0.79 $    &$ 44.01 $   &$5.81  $ &$ 1.05 $&$ 0.03  $\\
   HR 4158 &$ 6140$  &$ 4.22 $ &$-0.24 $   &$ -0.14 ~~$&$       $ &$ 0.78 $    &$ 40.67 $   &$5.71  $ &$ 1.05 $&$ 0.06  $\\
   HR 4395 &$ 6643$  &$ 3.98 $ &$-0.10 $   &$ 0.02  ~~$&$       $ &$ 0.25 $    &$ 22.80 $   &$5.08  $ &$ 1.50 $&$ 0.28  $\\
   HR 4529 &$ 6083$  &$ 4.04 $ &$0.16  $   &$ 0.17  ~~$&$ 0.15  $ &$ 0.58 $    &$ 23.49 $   &$6.24  $ &$ 1.33 $&$ 0.06  $\\
   HR 4540 &$ 6176$  &$ 4.14 $ &$0.13  $   &$ 0.17  ~~$&$ 0.05  $ &$ 0.57 $    &$ 91.74 $   &$3.59  $ &$ 1.33 $&$ 0.01  $\\
   HR 4657 &$ 6247$  &$ 4.38 $ &$-0.70 $   &$ -0.46 ~~$&$ -0.42 $ &$ 0.73 $    &$ 44.34 $   &$6.11  $ &$ 0.80 $&$ 0.07  $\\
   HR 4903 &$ 5953$  &$ 4.00 $ &$0.24  $   &$ 0.33  ~~$&$ 0.22  $ &$ 0.54 $    &$ 25.17 $   &$5.89  $ &$ 1.33 $&$ 0.15  $\\
\hline 
\end{tabular}
\label{result}
\end{table*}

\setcounter{table}{1}
\begin{table*}
 \caption{continued. $^{\star}$ indicates a derived upper limit in $\mathrm{[C/H]}$}
 \begin{tabular}{l|ccr|r|rr|rrrr}
 \hline
   Star   & $T_{\mathrm{eff}}$  & $\log g$ & $\mathrm{[Fe/H]}$  & $\mathrm{[C/H]}$ & $\mathrm{[O/H]}$ & $\log \tau_9$     & $\pi$ & $V$     &$\mathcal{M}/\mathcal{M}_{\odot}$&$\Delta\log g$\\
   Id.    & [K]   & [cgs] &	    &       &       & [Gyr]       & mas  & mag      & 		& [cgs] \\ 		
   \hline			   	                      
   HR 4989 & $6314$  & $4.25$  &$-0.28$    & $-0.10~~$ & $-0.25$ & $0.64$        &$ 55.49$    &$4.90 $  &$ 1.12 $&$ 0.07  $ \\
   HR 5338 & $6177$  & $3.94$  &$-0.11$    & $-0.03~~$ & $ 0.00$ & $0.48$        &$ 46.74$    &$4.07 $  &$ 1.33 $&$ 0.20  $ \\
   HR 5459 & $5720$  & $4.27$  &$0.15 $    & $0.22 ~~$ & $ 0.15$ & $0.62$        &$742.12$    &$-0.01$  &$ 1.00 $&$ 0.04  $ \\
   HR 5542 & $6001$  & $4.09$  &$0.13 $    & $0.15 ~~$ & $ 0.03$ & $0.69$        &$ 24.99$    &$6.30 $  &$ 1.33 $&$ 0.06  $ \\
   HR 5698 & $6341$  & $4.04$  &$0.01 $    & $0.14 ~~$ & $-0.02$ & $0.51$        &$ 29.27$    &$4.99 $  &$ 1.58 $&$ 0.22  $ \\
   HR 5723 & $6532$  & $3.93$  &$-0.13$    & $-0.02~~$ & $     $ & $0.41$        &$ 30.90$    &$4.92 $  &$ 1.41 $&$ 0.09  $ \\
   HR 5996 & $5831$  & $4.03$  &$0.23 $    & $0.17 ~~$ & $0.09 $ & $0.72$        &$ 34.60$    &$6.32 $  &$ 1.18 $&$ -0.18 $ \\
   HR 6189 & $6200$  & $3.98$  &$-0.56$    & $-0.48~~$ & $-0.48$ & $0.74$        &$ 23.02$    &$6.33 $  &$ 1.18 $&$ 0.00  $ \\
   HR 6243 & $6435$  & $3.80$  &$0.00 $    & $0.12 ~~$ & $     $ & $0.28$        &$ 27.04$    &$4.64 $  &$ 1.65 $&$ 0.14  $ \\
   HR 6409 & $6480$  & $3.94$  &$0.09 $    & $0.14 ~~$ & $-0.04$ & $0.38$        &$ 19.80$    &$5.53 $  &$ 1.58 $&$ 0.20  $ \\
   HR 6569 & $6675$  & $4.15$  &$-0.27$    & $-0.06~~$ & $     $ & $0.33$        &$ 45.72$    &$4.76 $  &$ 1.33 $&$ 0.02  $ \\
   HR 6649 & $6088$  & $4.24$  &$-0.34$    & $-0.21~~$ & $-0.29$ & $0.86$        &$ 30.55$    &$6.19 $  &$ 1.05 $&$ 0.16  $ \\
   HR 6907 & $6382$  & $4.15$  &$0.13 $    & $0.19 ~~$ & $0.01 $ & $0.35$        &$ 27.53$    &$5.90 $  &$ 1.41 $&$ 0.06  $ \\
   HR 7126 & $6655$  & $4.16$  &$0.21 $    & $0.10 ~~$ & $     $ & $0.19$        &$ 31.24$    &$5.56 $  &$ 1.50 $&$ -0.01 $ \\
   HR 7560 & $6146$  & $4.14$  &$0.09 $    & $0.02 ~~$ & $     $ & $0.63$        &$ 51.57$    &$5.12 $  &$ 1.25 $&$ -0.07 $ \\
   HR 7766 & $5964$  & $4.22$  &$-0.36$    & $-0.19~~$ & $-0.26$ & $0.94$        &$ 30.84$    &$6.26 $  &$ 1.00 $&$ 0.16  $ \\
   HR 7875 & $5991$  & $4.09$  &$-0.44$    & $-0.30~~$ & $-0.30$ & $0.87$        &$ 41.33$    &$5.11 $  &$ 1.12 $&$ 0.18  $ \\
   HR 8027 & $6285$  & $4.20$  &$-0.37$    & $-0.35~~$ & $-0.31$ & $0.73$        &$ 35.07$    &$5.76 $  &$ 1.12 $&$ 0.08  $ \\
   HR 8041 & $5813$  & $4.20$  &$0.11 $    & $0.27 ~~$ & $     $ & $0.89$        &$ 37.80$    &$6.21 $  &$ 1.05 $&$ 0.01  $ \\
   HR 8077 & $6166$  & $4.05$  &$-0.07$    & $0.07 ~~$ & $     $ & $0.62$        &$ 27.06$    &$5.94 $  &$ 1.18 $&$ 0.10  $ \\
   HR 8181 & $6139$  & $4.34$  &$-0.67$    & $-0.62~~$ & $-0.57$ & $0.96$        &$108.50$    &$4.21 $  &$ 0.80 $&$ 0.05  $ \\
   HR 8665 & $6228$  & $4.15$  &$-0.32$    & $-0.17~~$ & $     $ & $0.72$        &$ 61.54$    &$4.20 $  &$ 1.18 $&$ 0.16  $\\
   HR 8697 & $6288$  & $3.97$  &$-0.25$    & $-0.18~~$ & $     $ & $0.56$        &$ 37.25$    &$5.16 $  &$ 1.18 $&$ 0.01  $\\
   HR 8969 & $6255$  & $4.16$  &$-0.17$    & $-0.10~~$ & $     $ & $0.64$        &$ 72.51$    &$4.13 $  &$ 1.12 $&$ 0.07  $\\
   HD 6434 & $5813$  & $4.42$  &$-0.54$    & $-0.33~~$ & $-0.34$ & $1.10$        &$ 24.80$    &$7.72 $  &$ 0.90 $&$ 0.06  $\\
   HD 14938& $6164$  & $4.09$  &$-0.37$    & $-0.30~~$ & $-0.17$ & $0.67$        &$ 18.58$    &$7.20 $  &$ 1.05 $&$ 0.01  $\\
   HD 17548& $5977$  & $4.27$  &$-0.59$    & $-0.35^{\star}$ & $-0.48$ & $1.06$        &$ 18.90$    &$8.16 $  &$ 1.00 $&$ -0.13 $\\
   HD 25704& $5844$  & $4.43$  &$-0.85$    & $-0.68~~$ & $-0.57$ & $1.22$        &$ 19.02$    &$8.11 $  &$ 0.70 $&$ 0.25  $\\
   HD 51929& $5845$  & $4.28$  &$-0.64$    & $-0.58~~$ & $-0.43$ & $1.14$        &$ 26.58$    &$7.39 $  &$ 0.70 $&$ 0.09  $\\
   HD 69611& $5795$  & $4.29$  &$-0.58$    & $-0.40~~$ & $-0.28$ & $1.08$        &$ 20.46$    &$7.74 $  &$ 0.90 $&$ 0.10  $\\
   HD 78747& $5824$  & $4.45$  &$-0.64$    & $-0.56~~$ & $-0.38$ & $1.08$        &$ 25.16$    &$7.72 $  &$ 0.70 $&$ 0.19  $\\
   HD 130551&$6237$  & $4.25$  &$-0.62$    & $-0.36~~$ & $-0.46$ & $0.85$        &$ 20.94$    &$7.16 $  &$ 0.80 $&$ 0.18  $\\
   HD 165401&$5758$  & $4.31$  &$-0.47$    & $-0.34~~$ & $     $ & $1.18$        &$ 41.00$    &$6.80 $  &$ 0.90 $&$ -0.10 $\\
   HD 188815&$6181$  & $4.29$  &$-0.58$    & $-0.40~~$ & $-0.42$ & $0.86$        &$ 17.82$    &$7.47 $  &$ 1.05 $&$ 0.13  $\\
   HD 199289&$5894$  & $4.38$  &$-1.03$    & $-0.66~~$ & $-0.72$ & $1.26$        &$ 18.94$    &$8.28 $  &$ 0.70 $&$ 0.11  $\\
   HD 201891&$5867$  & $4.46$  &$-1.06$    & $-0.55^{\star}$ & $-0.33$ & $1.23$        &$ 28.26$    &$7.37 $  &$ 0.70 $&$ 0.22  $\\
   HD 205294&$6236$  & $4.01$  &$-0.36$    & $-0.24~~$ & $-0.28$ & $0.64$        &$ 18.78$    &$6.86 $  &$ 1.18 $&$ -0.02 $\\
   HD 210752&$5910$  & $4.25$  &$-0.64$    & $-0.50^{\star}$ & $-0.56$ & $1.15$        &$ 26.57$    &$7.44 $  &$ 0.70 $&$ 0.02  $\\
   HD 215257&$5983$  & $4.37$  &$-0.65$    & $-0.55~~$ & $-0.52$ & $1.06$        &$ 23.66$    &$7.41 $  &$ 0.70 $&$ 0.23  $\\
   HD 218504&$5945$  & $4.20$  &$-0.62$    & $-0.40~~$ & $-0.43$ & $1.03$        &$ 16.20$    &$8.11 $  &$ 0.70 $&$ 0.12  $\\ 
   \\
   \hline
   \end{tabular}
   \end{table*}

\par
A solar MARCS model was used to produce a synthetic spectrum of the Sun.
This spectrum was compared with a solar high-resolution flux spectrum 
(Kurucz et al. 1984)\nocite{kurucz84} in order to calibrate the oscillator strengths of some
larger and more important lines including our [C {\sc i}] line (see Table 
\ref{linelist}). The differential analysis is consequently performed relative to
the Sun.  
\par
The carbon line is supposed to be blended by a weak Fe {\sc i} line at
$8727.13\mbox{{\AA}}$. In our analysis, the $\log gf$ value 
($\log gf_{\mathrm{Fe~{\scriptscriptstyle I}}}=-3.93$) from VALD was used.
An upper limit of $\log gf_{\mathrm{Fe~{\scriptscriptstyle I}}}=-3.60$ can be
estimated from oscillator strengths of Fe {\sc i} lines of the same multiplet
observed in the solar spectrum (Lambert \& Ries 1977\nocite{lambert}). Using this upper limit, we find
that the contribution from the Fe {\sc i} line to the total equivalent width is
less than 15\% for the coldest and most iron rich of our programme stars, and
less than that for the others. 
Also, due to the differential analysis, the uncertainty in
the $\log gf_{\mathrm{Fe~{\scriptscriptstyle I}}}$ does not affect the derived
carbon abundances by more than 10\% for the hottest stars in the sample.
\par
In order to verify the assumption that the carbon line is responsible for most
of the equivalent width of the line at $8727\mbox{{\AA}}$, we studied the
thermal broadening by measuring the FWHM of the feature in the solar flux
spectrum (Kurucz et al. 1984). We determined the width of the 
combination of the instrumental profile and macroturbulence by fitting 
the neighbouring Si {\sc i} lines and next calculated the $8727\mbox{{\AA}}$
feature, alternatively assuming it to be due to [C {\sc i}]  and Fe {\sc i},
respectively. The FWHM of the line feature in the spectrum was measured to
$\mathrm{FWHM}_{\mathrm{8727~feature}}=0.20\mbox{{\AA}}$ while the carbon and 
the iron line in the calculated spectra were measured to
$\mathrm{FWHM}_{\mathrm{[C~{\scriptscriptstyle I}]}}=0.19\mbox{{\AA}}$ and 
$\mathrm{FWHM}_{\mathrm{Fe~{\scriptscriptstyle I}}}=0.15\mbox{{\AA}}$,
respectively. This indicates that the observed feature is primarily the [C {\sc i}] line.
\section{Resulting abundances}
The abundance determinations were made by using synthetic spectra, 
calculated for a model atmosphere, tailored for each star. The synthetic
spectra were convolved with macroturbulence, rotational 
and instrumental broadening profiles, in order to match the observed spectral
lines. 
In most cases the rotational convolution was not necessary. We based the
convolution parameters for each spectrum on the profiles of
the Si {\sc i} line at $8728.01\mbox{{\AA}}$ and the Si {\sc i} line at
$8742.45\mbox{{\AA}}$. There is, however, no unique way to convolve the spectra
to obtain the observed line profiles. By changing the parameters of the
convolution in different ways we could see that the derived carbon abundances
were not sensitive to the choice of convolution profile. The carbon abundance
was next changed until a good fit to the observed spectrum was provided.

The derived abundances are presented in Table \ref{result}. 
In the spectra for the stars HD 17548, HD 201891 and HD 210752 the carbon line
was too weak to allow an accurate measurement of the carbon abundance, so we
could only determine an upper limit. Those stars are excluded from further 
analysis.

\begin{table*}
  \caption{Effects on logarithmic abundances derived when changing the 
fundamental parameters of the model atmospheres. Three stars of different 
parameters are presented. The parameters ($T_{\mathrm{eff}}, \log g,  
\mathrm{[Fe/H]}, \xi_{\mathrm{t}}$) are, respectively:
$\mathrm{(5965,4.37,-0.82,1.39), (6480,3.94,0.09,2.36)}$ and
$\mathrm{(6517,4.01,-0.24,2.15)}$}
  \label{error}
  \begin{tabular}{l l l l l l l l l l l l l }
  \hline
  \noalign{\smallskip}
   & HR 3578  & & & HR 6409  &  & & HR 368  &  \\
   Change  & $\mathrm { \Delta [C/H] }$  & $\mathrm {\Delta [C/Fe] }$ & 
\,\,\,\,   & $\mathrm {\Delta [C/H] }$ & $\mathrm {\Delta [C/Fe] }$ & 
\,\,\,\,& $\mathrm{ \Delta [C/H] }$ & $\mathrm {\Delta [C/Fe] }$ \\
  \noalign{\smallskip}
  \hline
  \noalign{\smallskip}
  $\Delta T_{\mathrm{eff}}=+100\,\mbox{K}$ & $+0.03$ & $-0.03$ & & $+0.03$  &
$-0.03$ &  & $+0.04$ & $-0.02$ \\
  $\Delta \log g =+0.2 $ & $+0.07$ & $+0.08$ & & $+0.07$  & $+0.08$ &  & $+0.06$
& $+0.07$ \\
  $\mathrm { \Delta [Fe/H]=+0.1  }    $ & $+0.02$ & $+0.02$ & & $+0.01$  &
$+0.01$ &  & $+0.01$ & $+0.01$ \\

  \noalign{\smallskip}
  \hline
  \end{tabular}
\end{table*}
\section{Error analysis \label{error.Sec}}
\subsection{Errors in observed spectra and continuum location}
The largest statistical errors are introduced when measuring the abundance 
-- that is when determining the continuum and fitting the synthetic spectrum to 
the observed one. These errors vary from star to star depending on the quality
of the spectra.  The estimated mean error for the more metal poor half of the
stars is $\pm0.07~\mbox{dex}$ and for the metal rich half $\pm0.04~\mbox{dex}$.
\subsection{Errors in fundamental parameters}
The stars in our sample are chosen among the stars analysed 
in EAGLNT, where an extensive error analysis is also provided. 
A discussion on the model atmospheres, used also in this study, and the
derivation of the fundamental atmospheric parameters used for the model
atmospheres are presented in EAGLNT.
\par
The statistical errors of the effective temperature are estimated in EAGLNT to
be of the order of $50\, \mathrm{K} $. In addition to these random errors, 
systematic shifts in the temperature scale of $50-100\, \mathrm{K}$ should be
considered. This limit on systematic errors is consistent with the tests 
performed for some of the stars from EAGLNT by Tomkin et al. \cite*{tomkin}.
\par
The logarithm of the gravity has estimated statistical errors of  
$0.07~\mbox{dex}$, but with possible systematic errors of
$0.2~\mbox{dex}$ or even more, cf. 
the discussion below.
\par
The statistical errors in the model metallicity and in the microturbulence,
$\mathrm \xi_{\mathrm{t}}$, are estimated to be unimportant for the derived
carbon and iron abundances, cf. Andersson \& Edvardsson \cite*{andersson}.
\par
The propagation of these parameter 
errors to the carbon abundance estimate has been investigated by changing the
fundamental parameters with a representative amount and then running the
synthetic spectrum program to obtain the change in carbon abundance. This is
shown in Table \ref{error} for three stars, spanning the parameter space.
The statistical error in the carbon to hydrogen ratio due to the uncertainties
in the fundamental parameters is around $0.03~\mbox{dex}$. The systematic error
is larger, $0.08~\mbox{dex}$, mainly reflecting the significance of the errors
in $\log g$.

\subsection{Errors due to departures from LTE and oversimplified models}
The  carbon line being used here ($\mathrm {2p^1S - 2p^1D}$) is forbidden due to
the change of the orbital angular momentum quantum number by two units for
transitions between the two states. These low excitation states are the first
and second excited states of carbon. 
St\"urenburg \& Holweger \cite*{sturenburg} found for the Sun and Vega that 
non-LTE abundance effects of neutral carbon  increase with the strength of a
spectral line.
For the Sun, the departure coefficients of the levels included in the
[C {\sc i}] $8727\mbox{{\AA}}$ line  are negligible for all optical depths.
The lowest three levels of C {\sc i} are strongly coupled to the ground state
through collisions. In their model for Vega (A0 V) these low lying states show
strong under-population due to the stronger UV field, but are still strongly
coupled to each other.
Rentzsch-Holm \cite*{inga} showed in a non-LTE analysis of carbon that 
the non-LTE effects depend on equivalent widths of the lines.
In her $7000~\mbox{K}$ model the lowest three levels are in LTE all the way out
in the photosphere. 
Non-LTE effects increase with effective temperature.
It seems therefore unlikely that the [C {\sc i}] $8727\mbox{{\AA}}$ line will be
subject to non-LTE effects. Thus, no serious systematic errors are to be
expected due to our assumption of LTE. 
\par
Other systematic errors in the model atmospheres, e.g. in the 
temperature structure, errors in the basic assumptions of the modelling of the
atmospheres such as that of plane-parallel stratification and 
effects of unresolved binary stars, 
are in general terms discussed in EAGLNT. For the carbon abundances
derived here, these errors may be judged to be less significant than 
abundance determinations for most elements, in view of the
relatively low excitation energy of the line, its weakness and the fact that
it represents the dominating ionization stage of carbon. 
The error in the carbon abundance due to the errors in the determination of the
oscillator strength of the carbon line is only $\pm0.01~\mbox{dex}$
(Andersson \& Edvardsson 1994)\nocite{andersson}.

\section{Relations between abundances and comparison with other results}
\begin{figure}
 \resizebox{\hsize}{!}{\includegraphics{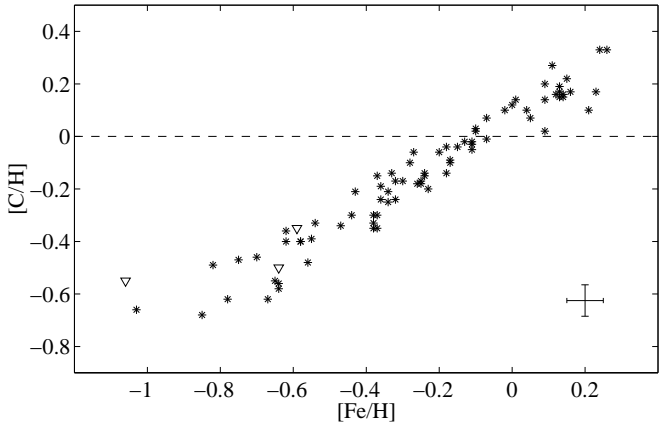}}
 \caption{$\mathrm{[C/H]}$ as a function of $\mathrm{[Fe/H]}$. The triangles
denote upper limits in $\mathrm{[C/H]}$ and will be excluded in the analysis.
A typical error bar is shown in the lower right corner}
 \label{CHvsFeH}
\end{figure}

In Fig. \ref{CHvsFeH} we show $\mathrm{[C/H]}$ vs. $\mathrm{[Fe/H]}$ for the
stars observed. Obviously there is a tight correlation. 
Assuming a linear model, the spread in the points is found to be consistent
with errors in $\mathrm{[Fe/H]}\sim0.05~\mbox{dex}$, as found by EAGLNT, and
with errors in [C/H] according to the random errors estimated in the discussion
above. 
The slope  $\mathrm{\Delta [C/H]/\Delta [Fe/H]}$ is $\mathrm{(0.85\pm 0.03)}$, 
with the error representing the statistical errors.
What are the effects of the systematic errors discussed above on this relation? 
A systematic trend is found when comparing the $\log g$-values derived from
Str\"omgren photometry (Edvardsson et al. 1993)\nocite{edvardsson} and the 
values obtained using
parallaxes measured by the Hipparcos satellite. For stars of low metallicities
an even lower abundance is derived -- that is, a steeper increase of
$\mathrm{[C/H]}$ vs. $\mathrm{[Fe/H]}$ is found -- if Hipparcos parallaxes
are used. The slope is in the latter case  $\mathrm{\Delta [C/H]/\Delta
[Fe/H]=}$ $\mathrm{(0.88\pm 0.03)}$.
\par
\begin{figure}
 \resizebox{\hsize}{!}{\includegraphics{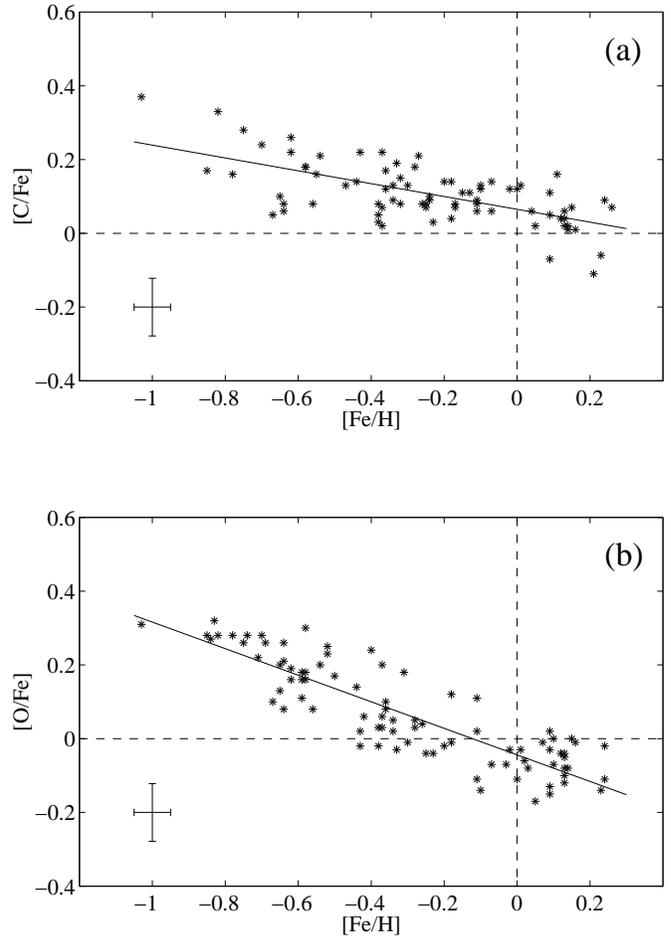}}
 \caption{\textbf{a} $\mathrm{[C/Fe]}$ as a function of $\mathrm{[Fe/H]}$.
The asterisks denote abundances of observed stars and the full line is the
determined slope given in Eq. 3. \textbf{b} $\mathrm{[O/Fe]}$ as
a function of $\mathrm{[Fe/H]}$ with data taken from Edvardsson et al. (1993).
The slope calculated from this data is given in Eq. 4. Typical error bars are
shown in the lower left corners
}   
 \label{CFeOFe}
\end{figure}

In Fig. \ref{CFeOFe}a, we have plotted the $\mathrm{[C/Fe]}$ ratios for the
observed stars as a function of $\mathrm{[Fe/H]}$. There is a well
defined relation with a small scatter around it:
\begin{equation}
\mathrm{[C/Fe]} = (-0.17\pm 0.03) \times \mathrm{[Fe/H]}  +  (0.065\pm 0.008).
\end{equation}
The standard deviation is $0.06$ which again is consistent with the estimated
random errors in carbon and iron abundances. Also in this case the effects on
the slope from possible systematic errors in $\log g$ are minute.
\par
The small scatter around the linear relations discussed here indicates that the  estimates made above of random errors in the carbon abundances are not
underestimated. The slope in the second relation, $-0.17\pm0.03$ is 
significantly less steep than that in the relation between [O/Fe] and [Fe/H],
\begin{equation}
\mathrm{[O/Fe]}=(-0.36\pm 0.02)\times \mathrm{[Fe/H]}  -  (0.044\pm 0.010),
\end{equation}
as obtained from the results of EAGLNT and 
also shown in Fig. \ref{CFeOFe}b. The slopes as
such reflect the slowly growing significance of iron-producing supernovae Type 
{\sc i}a during the evolution of the Galaxy, as compared with the rapid
production of oxygen by Supernovae Type {\sc ii} (see, e.g., EAGLNT). However,
the difference between these two slopes will be the major issue in our
discussion. 
\begin{figure}
\resizebox{\hsize}{!}{\includegraphics{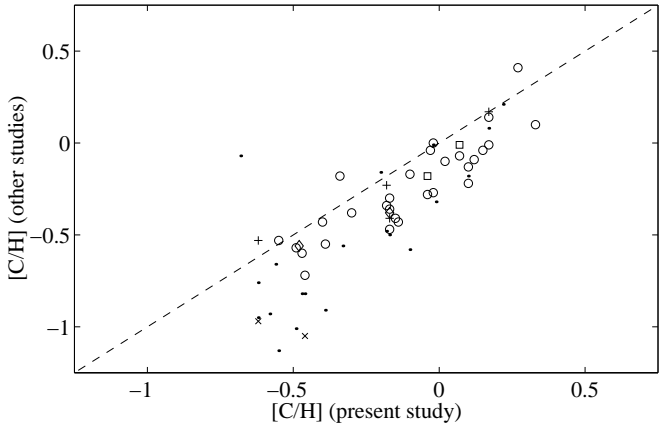}}
\caption{Comparison of derived $\mathrm{[C/H]}$ for the stars in common between
earlier studies and present study. Plusses ($+$) are from Clegg et al. (1981),
crosses ($\times$) are from Tomkin \& Lambert (1984), points ($\cdot$) are
from Laird (1985), squares ($\Box$) are from Friel \& Boesgaard (1990), diamonds
($\Diamond$) are from Friel \& Boesgaard (1992), and open circles ($\circ$) are 
from Tomkin et al. (1995).
The dashed line shows the loci of identical results
}
\label{jfr_C}
\end{figure}
\par
Carbon abundances of Galactic Disk stars have been determined by several 
other groups. 
We have 20 stars in common with the study of Laird \cite*{laird} and 29 in 
common with Tomkin et al. \cite*{tomkin}. Furthermore, we have a few stars in 
common with the studies of Friel \& Boesgaard (1990, 1992) and Tomkin \& 
Lambert \cite*{tomkin_lambert} and Clegg et al. \cite*{clegg}.
A comparison between the different results is shown
in Fig. \ref{jfr_C}.
\par
Andersson \& Edvardsson \cite*{andersson} found a slope in C/Fe, based on the
same line as we have used, however, for fewer stars and lower S/N, of $-0.2$.  
They also extracted data from previous analyses and separated abundances based
on separate features for comparison.
The abundances found by Laird \cite*{laird} from CH molecular lines show
a slope in the $\mathrm{[C/Fe]}$ vs. $\mathrm{[Fe/H]}$ of $-0.10$
if only stars with $\mathrm{[Fe/H]} \ge -1.0$ are considered.
The data show considerable scatter, however.
Clegg et al. \cite*{clegg} found a mean slope of $\approx -0.16 \pm 0.08$ from
highly excited C {\sc i} lines, the [C {\sc i}] $8727\mbox{{\AA}}$ line
and CH lines in F and G main-sequence stars. 
Their data demonstrate systematic differences between results from 
the different lines used:
The highly excited C {\sc i} and the [C {\sc i}] $8727\mbox{{\AA}}$ line give a
slope of $-0.18$ and the CH-based abundances give $-0.27$, while
a slope of $0.00$ was obtained from C$_2$ lines.
Friel \& Boesgaard (1990, 1992) \nocite{apj351} \nocite{apj387} show from 
highly excited C {\sc i} lines in field stars a slope of $-0.18$, while ,
more recently, Tomkin et al. \cite*{tomkin} derived from the C {\sc i} line at
$7100\mbox{{\AA}}$ of field disk stars a slope of $-0.28$.
\par
In Fig. \ref{jfr_C} we also note a small offset of our data; e.g. relative 
to those of Tomkin et al. (1995) we tend to obtain $\mathrm{[C/H]}$ ratios
greater by $0.1~\mbox{dex}$.
This may be due to the use of different carbon lines.
If our results were true they also indicate a tendency for the Sun to be
somewhat carbon poor or iron rich
(cf. also Fig. \ref{CHvsFeH}). This latter circumstance should be compared with
the possible tendency for the Sun to be somewhat poor also in other 
elements like Mg, Al and Si relative to Fe (Gustafsson 1998)\nocite{gust98}. 
In view of the possible systematic errors involved, e.g., in the gravity scale,
we do not consider our offsets from the identity line in
Fig. \ref{jfr_C} and our finding for the solar type stars to have non-solar
$\mathrm{[C/Fe]}$ ratios to be very significant. This, however, deserves further
study.
Our data show a smaller spread in the $\mathrm{[C/Fe]}$ vs. $\mathrm{[Fe/H]}$ 
relation than do 
the other studies. We ascribe this to the high quality of our data, and
the insensitivity of the $8727\mbox{{\AA}}$ [C {\sc i}] line to temperature. 

\section{Discussion}
\subsection{Possible sources of carbon}

As was mentioned in the Introduction a number of different production sites have
been suggested for carbon: supernovae, novae, Wolf Rayet stars, intermediate and
low mass stars in connection with the planetary-nebulae stages or even before by
superwinds at the end of the red-giant phase. Here, we shall discuss these
different possibilities in the light of our observations.  
\par
If carbon were produced essentially by SNe {\sc ii}, with yields relatively
independent of the initial metal abundances of the massive stars from which the
supernovae originate as is the case for oxygen, then one would expect the slope
in the two relations in Fig. 4 
to be the same. This is because the yield of carbon
produced by a generation of massive stars would scale with that of oxygen.
True enough, the $\mathrm{C/O}$ ratios of the nuclei produced and expelled by
SNe {\sc ii} of different mass may be different. E.g., from the tables of
Nomoto et al. \cite*{nomoto97np} we find $^{12}\mathrm{C}/^{16}\mathrm{O}$
ratios varying from $0.23$ (at $15 \mathcal{M}_{\odot}$), $0.21$
($18 \mathcal{M}_{\odot}$), $0.049$ ($25 \mathcal{M}_{\odot}$) and $0.016$
($40 \mathcal{M}_{\odot}$). An analogous
trend is found from the grid of supernovae models of Woosley \& Weaver 
\cite*{ww95apjs} (henceforth referred to as WW95)
although their $\mathrm{C/O}$ ratios are greater, reflecting  their lower rates
for the critical $\mathrm{^{12}C(\alpha ,\gamma )^{16}O}$ reaction.
However, the time scale difference between stars in the lower end of the mass
interval that produce SNe {\sc ii} and stars of very great masses is still much
smaller than the characteristic time scale for SNe {\sc i}a, producing iron.
Therefore, these different yields of supernovae of different mass are not
expected to lead to any significant slope differences in Fig. 4.
If the $\mathrm{C/O}$ ratios were dependent on the initial metallicity of the
star evolving to a supernova this would, however, lead to a difference.
In effect, if the most metal-poor stars would produce relatively more oxygen one
could at least qualitatively reproduce the slopes in Fig. \ref{CFeOFe}.
However, the SN {\sc ii} models with different metallicity do not predict any
substantial differences in the $\mathrm{C/O}$ ratios of the yields.
E.g., the models with solar initial abundances of WW95 have yields with
$\mathrm{O/C}=7.1$, when integrated with a mass distribution corresponding to a
Salpeter IMF, while those for 1/100 solar metallicity lead to
$\mathrm{O/C}=6.6$. 
Also note that the absolute amount of carbon produced
in the models mentioned, as compared with other SN {\sc ii} products, e.g., Mg,
is not sufficient to explain the abundance difference of a factor of 10 found,
e.g. in the Sun, cf. WW95 and Thielemann et al. \cite*{thiel}. 
Although there are uncertainties still in predicted carbon yields for
SNe {\sc ii} due to the uncertainty in the
$\mathrm{^{12}C(\alpha ,\gamma )^{16}O}$ rate and the details in the treatment
of convection, cf. WW95, the uncertainties do not seem large enough to admit an
increase of the carbon yields by a factor of 5. Obviously, another source of
carbon is needed.
\par
Supernovae of Type {\sc i}a, believed to produce most of the iron, do not
contribute significant amounts of carbon. E.g., the various models listed
by Nomoto et al. \cite*{nomoto97np} give carbon yields smaller than 10\% of
their iron yields. 
\par
Novae may also be ruled out as important sites for the production of carbon
since the frequency of such events and their mass loss are too low to enrich the
interstellar medium enough with common elements such
as $^{12}\mathrm{C}$ (Gehrz et al. 1998)\nocite{gehrz}.
\par
Massive stars may, however, contribute processed material in earlier phases, due
to their massive radiation-driven winds. These winds may then contain He and C
due to hydrogen and helium burning, but only smaller amounts of O and heavier
elements. Maeder \cite*{maeder} suggested yields from stars more massive than
$25 \mathcal{M}_{\odot}$ to be strongly dependent on metallicity, reflecting the
fact that more metal rich gas has a much greater cross section for radiation in
the ultraviolet.  E.g., a $40 \mathcal{M}_{\odot}$ solar composition model is
suggested to eject about 10\% of its mass as He and 10\% as C into 
the interstellar medium, while a corresponding model with 1/20 of the solar
metallicity hardly looses anything. These calculations have been detailed by
Portinari et al. \cite*{portinari} who also produce results for intermediate
metallicities. In their models, $\mathrm{C/O}$ of the winds is less than solar
for initial masses in the interval $12$ - $30 \mathcal{M}_{\odot}$.
Not until the mass is on the order of $40 \mathcal{M}_{\odot}$ or greater does
$\mathrm{C/O}$ become greater than one.
Although there is a general agreement between observed
and predicted relations for $\mathrm{[C/Fe]}$ vs. $\mathrm{[Fe/H]}$ and
$\mathrm{[O/Fe]}$ vs. $\mathrm{[Fe/H]}$, the fit is not perfect. E.g., for
carbon a bump with high carbon abundances is predicted around
$\mathrm{[Fe/H]}=-0.5$, with a steep slope down in $\mathrm{[C/Fe]}$ when
proceeding towards solar $\mathrm{[Fe/H]}$.  This is not seen in our
observations. On these grounds, however, the origin of carbon in massive stars
may certainly not be refuted. The substantial uncertainties in calculated
yields, both from supernovae and the massive winds in earlier stages of massive
stars, may well explain these deviations from observations.
\par
A more empirical estimate verifies the role that massive stars may
play through their winds. The number of WR stars within $3\mbox{kpc}$ from the
Sun, a volume for which the sample is thought to be complete, is about 60
(Conti 1988)\nocite{contiNASA}. Of these, about half are WC stars which then implies a
number of WC stars in the Galaxy on the order of 300. Typical mass-loss
rates from WC stars are $7 \cdot 10^{-5} \,\mathcal{M}_{\odot}$/year and
$\mathrm{C/He}$ mass ratios 
range in the interval $0.5$ - $3.3$ (de Freitas \& Machado 1988)\nocite{de}. 
This implies a total annual carbon contribution
to the ISM from the WC stars of about 
$0.01 \mathcal{M}_{\odot}$, which is a very considerable number. In model
calculations to be described below, we find a value fairly consistent with 
that, considering the uncertainties: 
with the yields of Portinari et al. \cite*{portinari} we obtain a total carbon
contribution from massive stellar winds of $0.0035 \mathcal{M}_{\odot}$ per
year, for a star formation rate of $1 \mathcal{M}_{\odot}$ per year in the
Galaxy, and a Salpeter IMF.  
\par
Stars of intermediate and low masses ($<10 \mathcal{M}_{\odot}$) may also be
significant contributors of carbon on the galactic scale, as was suggested by
Tinsley \cite*{tinIAU}, Sarmiento \& Peimbert \cite*{sarmiento} and
Timmes et al. \cite*{timmes}. The basic argument for these authors was that the
estimated contribution from supernovae and novae was not enough to account for
the observed carbon abundance. The theoretical yields of carbon from
intermediate and low mass stars are, however, highly uncertain, not the least
for the numerous low-mass stars. 
The pioneering calculation of yields by Renzini \& Voli \cite*{renz} were not
successful in predicting observed $\mathrm{C/O}$ and $\mathrm{N/O}$ ratios
for stars and planetary nebulae, nor could they reproduce the 
carbon stars of relatively low luminosities in the Magellanic clouds
(cf. {\it Note added in proofs} by these authors). These calculations are now
surpassed by those of
Marigo et al. (1996, 1998) who have published yields based on semi-analytical
models of AGB stars with mass loss, with several free parameters to describe
dredge-up of processed material. They find significant contributions of carbon
from stars in the mass interval $1.5$ - $4 \mathcal{M}_{\odot}$.
Forestini \& Charbonnel \cite*{forestini} have calculated evolutionary models
to the tip of the AGB with pulsations included and found significant
contributions of carbon for the lowest masses, amounting to a few percent of a
solar mass for stars around $2$ - $2.5 \mathcal{M}_{\odot}$. Both these papers
suggest greater yields of carbon for low metallicities (Fig. \ref{p_corr}).
Empirically, the finding that 50\% of the planetary nebulae are carbon rich
(with $\mathrm{C/O}\geq 1$), (Zuckerman \& Aller 1986\nocite{zuckermann}; 
Rola \& Stasinska 1994\nocite{rola}) and a PN birth rate of 1 per year in the
Galaxy (Pottasch 1992)\nocite{pottasch} and a PN mass of $0.3 
\mathcal{M}_{\odot}$, suggests a
carbon contribution of about $0.003 \mathcal{M}_{\odot}$ per year to the ISM.
With such a contribution, the PNe could be a major source of carbon. One may ask
whether mass loss from carbon stars, i.e. before the PN stage,  also could be a
major source. Adopting typical mass-loss rates 
(see Olofsson et al. 1993, 1996\nocite{olofsson93}\nocite{olofsson96}; 
Jura \& Kleinmann 1989\nocite{jura}), $\mathrm{C/O}$ ratios (Lambert et al. 
1986)\nocite{lambert86apjs} and
carbon-star frequencies we find a total carbon contribution of $2\cdot 10^{-4}$
$\mathcal{M}_{\odot}/\mathrm{year}$, which should not be very significant in
the global carbon budget of the Galaxy.
We note also that Prantzos et al. \cite*{prantzos}, argue that the time scale
for the halo phase of about $1$ - $2~\mbox{Gyrs}$ admits the contribution of
carbon by intermediate mass stars to the initial abundance of the Galactic Disk,
as predicted from the yields of Renzini \& Voli \cite*{renz}, but
that the lack of a peak in the $\mathrm{C/O}$ ratio around
$\mathrm{[Fe/H]}=-1.0$ suggests other production sites. They conclude from their
model calculations that an origin in massive winds as suggested by
Maeder \cite*{maeder} is most probable.

\subsection{Tests of possible carbon sources}
One may now ask which of the two major remaining carbon sources  -- massive
stars through massive stellar winds, presumably in the Wolf Rayet WC stage,  
or the contribution from intermediate/low mass stars through planetary nebulae
-- is the most significant source. These two different sources could reveal 
themselves in different strengths of the correlations between carbon abundance 
and metallicity, and age, respectively. 
The WC origin of carbon should be expected to lead to a stronger
correlation of the carbon abundance with metallicity, 
while if the age is the dominating factor, determining
the carbon abundance as in the planetary nebulae case, one would expect
a stronger correlation between $\mathrm{[C/H]}$ and age. 
\begin{figure}
 \resizebox{\hsize}{!}{\includegraphics{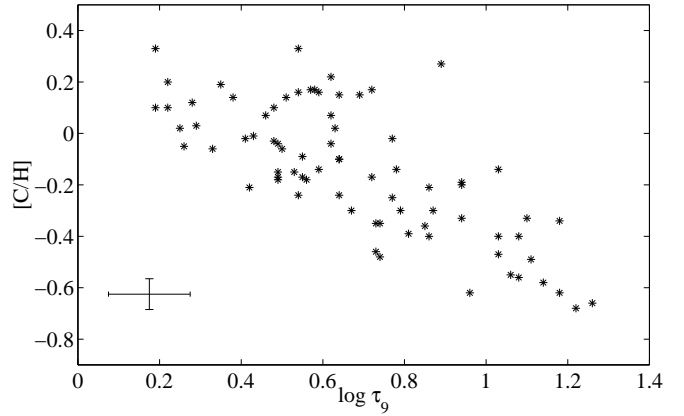}}
 \caption{$\mathrm{[C/H]}$ as a function of $\log\tau_9$. The scatter in this
 relation is much greater than in the $\mathrm{[C/H]}$ vs. $\mathrm{[Fe/H]}$
 relation (Fig. 3), partly due to errors in the age estimates}   
 \label{CHvslogtau9}
\end{figure}
\par
In contrast to the strong correlation between the carbon abundance and
metallicity, as illustrated in Fig. \ref{CHvsFeH}, a much greater scatter is
found in the $\mathrm{[C/H]}$ vs. stellar age relation (Fig. \ref{CHvslogtau9}).
A similarly great scatter is found in the [Fe/H]-age diagramme
(see EAGLNT, their Fig. 14a). This could indicate a cosmic spread in [Fe/H] at a
given age, which then indicates that $\mathrm{[C/H]}$ is more tightly related to
metallicity than to age. Another possibility is that the scatter in 
Fig. \ref{CHvslogtau9} is due to random errors in age. If so, these errors must be
as large as $0.2~\mbox{dex}$ or around 60\%.
The estimated errors in the ages are more likely on the order of 25\%.
Therefore, it seems likely that there is a real spread in the metallicity-age
relation and that the primary dependence is on metallicity. This argument is
weakened, however, by the possibility that the spread in the  
$\mathrm{[Fe/H]-age}$ relation
is due to the diffusion of stars formed at different distances from the Galactic
centre into orbits close to the solar one (Wielen et al. 1996)\nocite{wielen}.
\par
\begin{figure}
 \resizebox{\hsize}{!}{\includegraphics{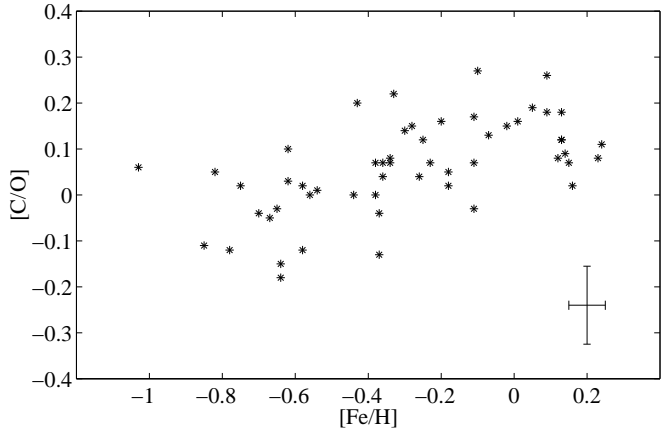}}
 \caption{$\mathrm{[C/O]}$ as a function of $\mathrm{[Fe/H]}$. We can see a
gradual increase in $\mathrm{[C/O]}$ with $\mathrm{[Fe/H]}$. This observed
enhancement of carbon to the ISM could be a result of carbon production in
massive stars with metallicity dependent yields or intermediate- and 
low-mass stars, or possibly a mixture  
}   
 \label{COvsFeH}
\end{figure}

We here suggest another test, built on the following argument: If carbon is
produced in massive stars, but with metal-abundance dependent yields due to the
radiatively driven winds, one would expect a gradual increase of the
$\mathrm{C/O}$ ratio with $\mathrm{[Fe/H]}$. This is also observed in the Disk,
(Fig. \ref{COvsFeH}). Similarly, if less massive stars produce carbon and
expel it as PNe, one again expects a gradual increase of $\mathrm{C/O}$ in the
Disk, reflecting the difference in time scale between the carbon producing stars
and the massive stars producing oxygen. 
\par
Our test is built on the fact that these two factors behind the $\mathrm{C/O}$
increase -- metallicity and time-scale -- are interrelated in different ways in
different galaxies. Thus, irregular galaxies, like the Magellanic Clouds, have a
lower metallicity ($\mathrm{[Fe/H]_{irr}}$), while the time scale for a more or
less continuous star formation history may be of the same order of magnitude as
for the Galactic Disk. For these irregulars we therefore expect the massive star
contribution of carbon relative to oxygen to be be comparable to that from
massive stars in the Disk, when its metal-abundance was $\mathrm{[Fe/H]_{irr}}$.
Thus the same slope should be delineated in the
$\mathrm{[C/O]}$-$\mathrm{[Fe/H]}$ diagram by the irregulars and the Galactic
Disk stars. For the contribution of carbon from intermediate- and low-mass 
stars, however, we
expect a smaller contribution in the Disk when the metal abundance was
$\mathrm{[Fe/H]_{irr}}$ if the latter is small enough ($\leq -0.5$), since time
then had still not admitted these stars to contribute much. In the more
slowly evolving irregulars, however, these contributions are fully developed.
Thus, we predict a greater slope in the $\mathrm{[C/O]}$-$\mathrm{[Fe/H]}$
diagramme for the Galactic Disk stars if intermediate- and low-mass stars are 
more significant as carbon sources.
\par
It should be noted that Edmunds \& Pagel \cite*{edmunds} pointed out that,
provided that nitrogen would be formed as a primary element in low-mass stars,
the N/O-O/H diagramme could be used for dating galaxy populations. Here, we 
use a similar idea, for another purpose.    

\subsection{Models of carbon synthesis in galaxies}
We shall now illustrate our discussion by using models of Galactic chemical
evolution. Such models must, however, be used with care -- in particular
when compared quantitatively to real observations -- 
since they are marred by the uncertainties in the in-going data, in particular
in yields. Also, the uncertainties in the basic assumptions made in the models
are significant.  This is illustrated in Fig. 3 of Carigi \cite*{carigi} where a
number of different models from the literature are intercompared. 
\par
Our chemical evolution model is based on the analytical
model of the local outer Disk, developed by Pagel \& Tautvai\v{s}ien\.{e} 
\cite*{pagel_taut}. This model uses the inflow formalism of
Clayton \cite*{clayton} and the delayed production approximation introduced by
Pagel \cite*{pagel}.
\par
The differential equations describing the chemical evolution are:

\begin{equation}
\frac{\mathrm{d}Z_1}{\mathrm{d}u}+\frac{F}{\omega g}Z_1=y_1(Z(u))
\end{equation}
\begin{equation}
\frac{\mathrm{d}Z_2}{\mathrm{d}u}+\frac{F}{\omega g}Z_2=0\qquad\qquad
\mathrm{if} u<\omega\Delta;
\end{equation}
\begin{equation}
\frac{\mathrm{d}Z_2}{\mathrm{d}u}+\frac{F}
{\omega g}Z_2=y_2(Z(u))\frac{g(u-\omega\Delta)}{g(u)}
\qquad \mathrm{if} u\ge\omega\Delta,
\end{equation}
where the total mass fraction of a species, $Z_i=Z_1+Z_2$, is the sum of the
mass fractions from instantaneous and delayed production. Here, $Z$ is the
overall metallicity (per mass). The time-like integration variable $u$, is
defined as
\begin{equation}
u\equiv\int_{0}^{t}\omega(t')\mathrm{d}t',
\end{equation}
where $\omega$ ($=0.3~\mbox{Gyr$^{-1}$}$) is the transition probability for gas
to change into stars in unit time at time $t$. The function describing the
inflow is of the form
\begin{equation}
\frac{F}{\omega g}=\frac{k}{u+u_0}
\end{equation}
and can be chosen to be switched on at a specific time $u_1$.
$F$ is the mass inflow rate, while the parameters $k$ and $u_0$ are choosen to 
fix the ratio of final to initial mass of the system. In this model, $u_1=0.14$,
$k=3$ and $u_0=1.3$, as in Pagel \& Tautvai\v{s}ien\.{e} 
\cite*{pagel_taut}. Note that before the inflow begins, the second term on
the left hand side of Eqs. 5, 6 and 7 is identically zero. 
\par
Eventually, the evolution of the gas mass, $g(u)$, is 
given by
\begin{equation}
\frac{dg}{du}+g=\frac{F}{\omega}.
\end{equation}
\par
All parameter values describing the chemical evolution are the same 
as those chosen by Pagel \& Tautvai\v{s}ien\.{e}
except for their choice of constant net yields. We 
explore the effects of 
metallicity dependent yields $p_{\mathrm{iM}}(Z(u))$, where $Z(u)$ is the
overall metallicity. The net yields ($y_{\mathrm{i}}$ in Eqs. 5 and 7) are
mass-integrated stellar yields divided by the lock-up fraction, which is treated
in the instantaneous recycling approximation. 
\par

\begin{figure}
 \resizebox{\hsize}{!}{\includegraphics{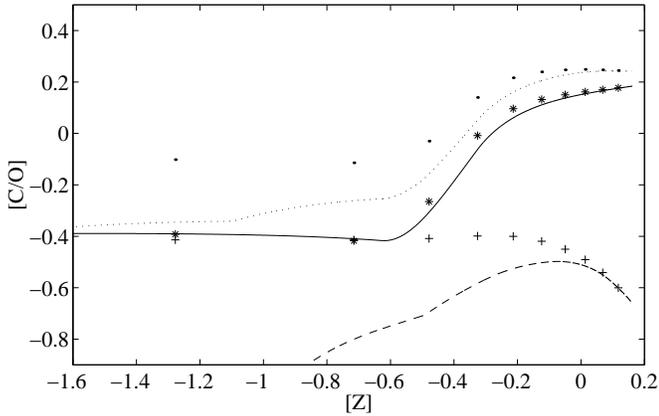}}
 \caption{Synthetic $\mathrm{[C/O]}$ ratio of the local disk and dwarf irregular
galaxies derived from a chemical evolution model using metallicity dependent
yields (Portinari et al. 1998). The full line shows the $\mathrm{[C/O]}$ ratio
of the Galactic Disk with carbon produced by high mass stars, the dashed line
shows the Disk with products from intermediate and low mass stars, the dotted
line is the combined $\mathrm{[C/O]}$ ratio with carbon from high, intermediate
and low mass stars in the Disk. The asterisks ($\ast$) represent
dwarf irregular galaxy models with carbon from high mass stars, the
plusses ($+$) are dwarf irregular galaxies with carbon from intermediate and
low mass stars and the points ($\cdot$) represent dwarf irregular galaxies
with carbon produced by high, intermediate and low mass stars. 
The sharp upturn in $\mathrm{[C/O]}$ at $[Z]=-0.6$ is a result of the
highly increasing theoretical carbon yield of high mass stars with increasing
metallicity}
 \label{model}
\end{figure}
\par
We have used carbon and oxygen yields of Portinari et al. \cite*{portinari} 
for high mass stars and carbon yields of Marigo et al. (1996, 1998) 
\nocite{marigo96} \nocite{marigo98} and Forestini \& Charbonnel 
\cite*{forestini} for intermediate and low mass stars. With an adopted 
linear time-metallicity relation normalized to the solar value, i.e. 
$Z=0.0055u$, we can calculate the evolution of
the $\mathrm{[C/O]}$ ratio with $Z$ (Fig. 
\ref{model}). Oxygen 
was assumed to be recycled instantaneously while carbon was divided into an 
instantaneous part (representing carbon production by high-mass stars)
and several delayed production parts with different time 
delays $\omega\Delta$, from $\omega\Delta=0.015$ to $\omega\Delta=1.000$, 
corresponding to carbon production by stars with masses from $6$ - $9 M_{\odot}$
to $1$ - $2 M_{\odot}$.
\par 
The dwarf irregulars were simulated like local Disk systems (i.e. same
parameters as the Disk), however,
with different slopes of the time-metallicity relation 
in order to mimic the slower metal production of heavy elements 
leading to a lower present metallicity. All systems were evolved for
$15~\mbox{Gyrs}$ ($u=4.5$) and thereafter the $\mathrm{[C/O]}$ ratio was read
off for comparison with observations.
\par
The results of the simulations are shown in Fig. \ref{model}. It is seen that
the $\mathrm{C/O}$ ratio for the Galactic Disk, if the carbon is produced 
by massive stars, shows a gradual
and rather steep increase with Disk star metallicities, reflecting the
metal-abundance sensitivity of the radiatively driven winds. The 
models representing dwarf irregulars line up nicely along the Disk curve.
When carbon production from intermediate and low mass stars is added,
the dwarf irregulars are elevated above the $\mathrm{C/O}$ ratios of the Disk,
in particular for the low metallicity systems. Thus, in this case the 
dwarf irregulars should delineate a relation with smaller slope in the 
diagramme, as was argued on qualitative grounds above. 
In the extreme case (lower part of the figure) when only
intermediate- and low-mass stars contribute the slope difference is very
pronounced.

\begin{figure}
 \resizebox{\hsize}{!}{\includegraphics{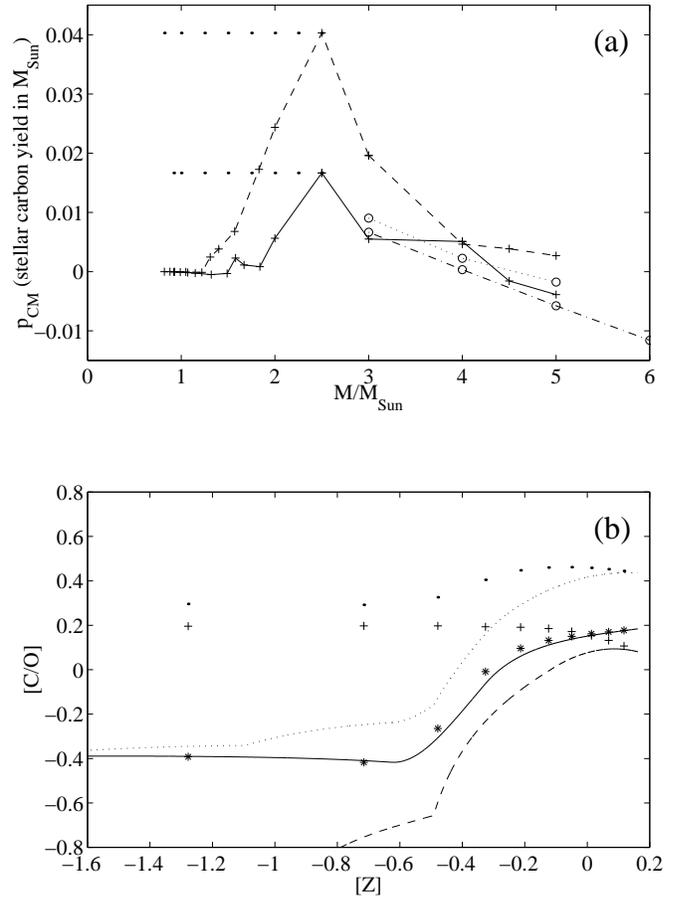}}
 \caption{\textbf{a} Theoretical stellar carbon yields
(in $\mathcal{M}_{\odot}$) for intermediate and low mass stars. The full line is
the yield at $Z=0.02$ and the dashed line is the yield at $Z=0.008$ from
Marigo et al. (1996, 1998).
The dash-dotted and the dotted lines show the yield at $Z=0.02$ and $Z=0.005$ 
respectively, calculated by Forestini \& Charbonnel (1997).  
Alternative increased carbon yields for low mass stars ($0.95$ - $2.5 
\mathcal{M}_{\odot}$) are indicated as dotted horizontal lines. 
\textbf{b} Synthetic $\mathrm{[C/O]}$ ratios in the Galactic Disk with the
increased carbon yields for low mass stars. Symbols as in Fig. 8}
 \label{p_corr}
\end{figure}
\par
Since the relative contribution to the production of carbon of stars of
different masses is highly uncertain we have ad hoc increased the yields of
Marigo et al. for the stars with masses below $2.5 \mathcal{M}_{\odot}$
according to Fig. \ref{p_corr}a. Such a change will guarantee that the frequency
of planetary nebulae with $\mathrm{C/O}\ge1$ gets as high (or even higer) than
observed, and that the absolute amount of carbon produced is as high as
observed. This is illustrated in Fig. \ref{p_corr}b. Again we notice the very
pronounced slope difference between the Disk star relation and that of dwarf 
irregulars.
 
\subsection{Results of the tests}
Before we discuss the results of the test it should be noted that
our observations of stars in the Solar neighbourhood may not map the chemical
evolution all the way back to the time of formation of the Disk.
Short lived intermediate-mass stars could have enriched the ISM, provided that 
they are sites of high carbon production. This is
in clear contradiction, though, to the theoretical yield calculations (Fig. 
\ref{p_corr}a) which indicate that the intermediate-mass stars in the upper mass region do not produce carbon in large amounts. They are also less numerous in comparison to stars of lower mass which further limits their relative contribution to the ISM. Moreover, if intermediate-mass stars were the main origin of carbon in the Disk in general, the carbon yield has to be metal 
dependent as for the massive stars, again in contradiction to the 
theoretical yields which rather decrease with increasing metallicity. However, 
the yields are uncertain and observationally, we are not able to exclude them as carbon producers of some importance.
\begin{figure}
 \resizebox{\hsize}{!}{\includegraphics{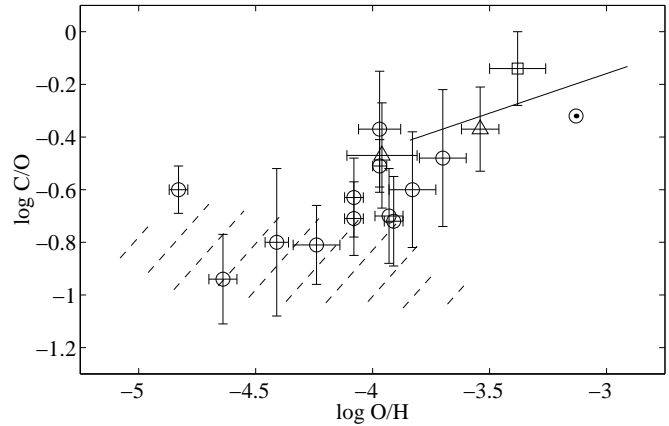}}
 \caption{Observed $\mathrm{C/O}$ ratios in various metal-poor galaxies. 
Open circles ($\circ$) represent data for H {\sc ii} regions in different
galaxies from Kobulnicky \& Skillman (1998) and references therein, the square
($\Box$) is the mean $\log \mathrm{C/O}$ ratio calculated 
from several observations of the Orion nebula. The triangles ($\triangle$) are
$\mathrm{C/O}$ ratios deduced from supergiant stars in LMC and SMC by Hill
et al. (1995, 1997). The full line indicates the calculated slope 
of the Galactic Disk stars, derived from observed carbon and oxygen abundances.
The hatched area in the lower part of the figure indicates the location of halo
dwarf stars, studied by Tomkin et al. (1992). The location in the horizontal
direction of these stars is relatively uncertain due to temperature
uncertainties. The Sun is also indicated in the 
figure as a $\odot$. The slope of the Disk stars is not steeper than
the slope of the dwarf irregulars, indicating that low-mass
stars are not significant carbon contributors relative to metal-rich high-mass 
stars. The role of the short lived intermediate-mass stars should be further investigated}
 \label{DGvsDS}
\end{figure}
\par
In order to carry out the test empirically we first use the observations of
massive supergiants in the Magellanic Clouds. The determinations for supergiants
in the SMC (Hill et al. 1997)\nocite{hill97} and LMC 
(Hill et al. 1995)\nocite{hill95} are plotted in Fig. \ref{DGvsDS}.
The relation for the Disk stars was obtained by combining Eqs. 3 and 4. 
From this we find that the stars in the Magellanic Clouds seem to take positions
in the diagramme characteristic of the Galactic Disk stars
which does not support the hypothesis of C being made in intermediate- or 
low-mass stars.
However, the supergiant CNO abundances may be affected by dredge-up of CNO
processed material, which makes the value of this test uncertain.
\par
In Fig. \ref{DGvsDS} we have also plotted the location of H {\sc ii} regions
in various metal-poor galaxies according to
Kobulnicky \& Skillman \cite*{kob_skill} and references therein and the locus of halo dwarf stars according to Tomkin et al. \cite*{tomkin_lemke92}.
There seems to be a tendency for the halo stars to have low C/O ratios as
compared with Disk stars. This may, however, be due to systematic errors
is the analysis of the halo stars, or possibly of the Disk stars (the latter is 
suggested by the deviating position of the sun in Fig. \ref{DGvsDS}),
but the suggested tendency is worth further investigation.
\par
In Fig. \ref{DGvsDS}
we find the galaxies to define a slope that seems as steep or even steeper than 
that of the Galactic Disk stars. Although adopting an [O/Fe] slope of 
$-0.40$ (cf. Eq. 4 and Fig. \ref{CFeOFe}b) the picture does not alter 
significantly. Actually, if we drop the mathematically derived equations (Eqs.
3 and 4), a somewhat higher value of both the [O/Fe]- 
and [C/Fe] slope may be argued for just by inspecting Fig. \ref{CFeOFe} by eye.
The C/O slope would then again decrease towards the adopted value (Fig. 
\ref{DGvsDS}). Even if the slope of the Disk stars is further steepened when 
the halo stars are added, the locii
of the metal-poor galaxies do not seem compatible with the hypothesis that
carbon has been formed mainly in stars of lower mass. The most direct conclusion from
this would thus be that the production of carbon occurs at a time-scale
characteristic of high-mass or possibly intermediate-mass stars of relatively
high mass, but that the yields are metal-abundance dependent, just as is
expected from massive stellar radiatively driven winds, e.g. from WC stars. 
\par
There may be several arguments against a test of this character. One may be the
possibility that carbon, and oxygen as well as iron, could be severely depleted
by grain formation in the interstellar medium of the irregular galaxies.
A depletion independent of metallicity for carbon, but increasing with
metallicity for oxygen, seems possible and would qualitatively lead to the
effect observed. It is, however, questionable whether this effect could be
as great as required to explain the full effect or to fully compensate for the
effect due to carbon formation by low-mass stars. 

Another possibility may be that the remnants from the oxygen producing SNe are
lost from the irregulars, and more so later in their evolution when the gas
density in the galaxy and its surroundings is less, while the slower carbon-rich
winds from the PNe are mostly retained. This would, however, diminish the slope
of the irregulars in the diagramme, which would then strengthen our
arguments for the high-mass stars as main contributors of carbon. 

The dwarf irregular galaxies could be systems of
much smaller age than the Galactic Disk. E.g., even if Garnett et al. 
\cite*{garnet_skillman97} have pointed out that a population of stars older than
the present star burst is probable in I Zw 18, the mean age of stars
in this system could still be much younger than the Galactic Disk. However,
at least in the Large Magellanic Cloud there is strong evidence for a dominating
population of an age comparable to that of the Galactic Disk 
(Ardeberg et al. 1997)\nocite{ardeberg97}. 
\par
Finally, if
the main source of carbon is high-mass stars one has to explain why planetary
nebulae, found above to provide as much as $0.003 \mathcal{M}_{\odot}$ of carbon
per year to the Galaxy today, do not constitute a major carbon source.
The statistics of PNe must comprize low-mass stars ($\le2 \mathcal{M}_{\odot}$)
as a major group; in view of the great PN birth rate. Therefore, if the masses,
the birth rate or the carbon abundances of the PNe are not seriously
overestimated, one has to advocate metal-abundance sensitive yields for these,
such that the amount of carbon produced be greater the higher the
metal-abundance. This is contrary to the tendency found in model calculations
both by Marigo et al. (1996, 1998) and by
Forestini \& Charbonnel \cite*{forestini}.

\section{Conclusions}

We have determined accurate carbon abundances for a set of 80 Galactic 
Disk stars. We find the resulting $\mathrm{[C/H]}$ values to correlate very 
well with the iron abundances $\mathrm{[Fe/H]}$, with a slope, however, that is
smaller than unity such that
$\mathrm {\Delta [C/Fe]/ \Delta [Fe/H]}=-0.17\pm 0.03$. 
From this, and the corresponding variation of $\mathrm{[C/O]}$,
we conclude by comparing to dwarf irregular galaxies of 
different metallicity that 
carbon is predominantly produced in the Galactic Disk by massive stars,
ejecting carbon in radiatively driven massive winds, in particular during
their Wolf Rayet stage. Estimates of yields from massive stars support
this conclusion, even if yields from intermediate- and low-mass stars 
also suggest those to be of some significance as sources of carbon. A particular
problem is raised by the estimates that about half the planetary nebulae
may be carbon rich (with $\mathrm{C/O}\ge1$); if so one has to explain where this 
excess carbon has gone. In particular, why is it not visible 
as a steeper slope for the Disk 
stars than for the dwarf galaxies in the $\mathrm{[C/O]}$ -- $\mathrm{[O/H]}$
diagramme? 
\par
The systematic study of carbon abundances of WR stars, carbon stars and
planetary nebulae of different progenitor mass and metallicity in
the Galaxy and neighbouring galaxies should be persued, in order to
establish more reliable empirical yields for high-, intermediate- and low-mass
stars. These abundances should also be related to the abundances of nitrogen and
the s elements which are thought to be formed in intermediate- and low-mass
stars.  A test similar to that applied here with metal-poor galaxies should be
possible and rewarding by using radial gradients in C/O and O/H measured for
disk galaxies. Also, isotopic characteristics of WR stars in solar-system
material, e.g. as presolar 
grains in meteorites, should be further looked for. At present that
situation is not clear; according to the recent review by Arnould et al.
(1997)\nocite{arnould} there is to date no clear and unambiguous signature of
the existence in meteorites of presolar grains of WR origin identified in the 
laboratory, while grains of AGB star origin have most probably been identified
(Anders \& Zinner 1993)\nocite{anders}.

\begin{acknowledgements}
The main contents of this study were presented at a symposium at Nordita,
Copenhagen, in May 1998 to celebrate Bernard Pagel's very significant 
contributions to Nordita and to astronomy in the Nordic countries. We 
dedicate this paper to him for his lasting inspiration and many fruitful
discussions. 
\par
Sofia Feltzing helped in carrying out some of the observations. 
Patrick de Laverny and Bertrand Plez are thanked for providing line data on CN.
We also thank the referee, Evan Skillman, for valuable comments on the
manuscript. BE and BG acknowledge support from the Swedish Natural Sciences Research Council (NFR) and NR acknowledges support from the Swedish National Space Board. 

\end{acknowledgements} 

\bibliographystyle{astron}
\bibliography{mnemonic,ref_C}

\end{document}